\documentclass[12pt]{article}

\usepackage{setspace}
\usepackage{amssymb}
\usepackage{color}
\usepackage{graphicx}
\usepackage{subfigure}
\usepackage{vmargin}
\usepackage{pstricks}
\usepackage{mathrsfs}
\usepackage{amsmath}
\usepackage{pst-plot}
\usepackage{pst-node}
\usepackage{verbatim}
\usepackage{authblk}

\newcommand{\etal}{\textit{et al.}}

\begin{document}

\title{A Truncation Error Estimation Scheme for the Finite Volume Method on Unstructured Meshes}

\author{A. R. Baserinia}

\affil{Plethora Corporation, San Francisco, CA 94124, United States}

\maketitle

\section{Abstract}
This work is an attempt to develop an approximate scheme for estimating the 
volume-based truncation errors in the finite volume analysis of laminar flows. The volume-based truncation error
is the net flow error across the faces of a control volume.
Unfortunately, truncation error is not a natural outcome of the finite volume solution and
needs to be estimated separately. Previous works in the literature estimate 
truncation error using either higher order interpolation schemes, higher order discretization 
schemes, or neglected terms in the discretization scheme. The first two approaches 
become complicated on general unstructured meshes  and the third approach provides 
inaccurate results.
This work proposes a truncation error estimation scheme, which is based on the third approach, 
but provides more accurate results compared to the existing results in the literature. 
The potential application of such a truncation error estimation scheme is in mesh adaptation.

\section{Introduction}


In numerical simulation of fluid flows, solution error is inevitable. This error is primarily due
to the discretization of the governing equations. In other words, the discrete solution
obtained from a numerical simulation is slightly different from the exact solution
of the governing equations. Therefore substituting the numerical solution into the exact governing 
equations results in an apparent source term, called the truncation error. 

The main application of solution truncation error is in CFD problems in which the assessment
and reduction of the solution error is crucial. In general, there are two approaches that
adopt the concept of truncation error to reduce the discretization error. The first approach is the use of
the solution truncation error as a source term in the governing equations to construct a higher 
order discretization scheme~\cite{celik:2017, ervin:1989, pierce:2004, yan:2017}. The second approach is the use of the solution
truncation error as an indicator for driving a mesh adaptation process~\cite{reuss:2003, roe:2002, syrakos:2012, zhang:2000}. 
This work concentrates on a truncation error estimation scheme that could potentially be used for the second purpose.

In the application of the solution truncation error as an indicator for mesh adaptation, we are interested in having
a reasonable estimate of the distribution of the solution truncation error. However, this estimate does not need
to be highly accurate. The reason is that the truncation error estimate will eventually  be used for determining
the characteristic sizes of mesh elements. In practice, we do not have complete control over these
characteristic size. Therefore even an approximate truncation error estimator that correctly predicts the overall trend
of truncation error distribution is adequate. The objective of this paper is
to present a simple and straightforward truncation error estimator that can be used in mesh adaptation applications.

\section{Conservation Equations and FVM Discretization}
In this work, we focus on the conservation equations for two-dimensional steady-state laminar
incompressible isothermal flows of constant-property Newtonian fluids in the absence of
gravity, which are governed by the mass and momentum equations:
\begin{gather}
  \rho \nabla \cdot {\bf v} = 0 \label{eq:1} \\
  \nabla ( \rho {\bf v}\otimes{\bf v} ) = -\nabla p + \nabla \cdot 
	\Big[ \mu \left( \nabla {\bf v} + \nabla {\bf v}^{\rm T} \right) \Big]  \label{eq:2} 
\end{gather}
where ${\bf v}=(u,v)$ is the flow velocity vector, $p$ is the pressure, 
$\rho$ is the density, $\mu$ is the viscosity, and the superscript $[\;]^{\rm T}$ is the transpose operator.
Note that the mass equation~\eqref{eq:1} is scalar while the momentum 
equation~\eqref{eq:2} is vectorial. Therefore the entire system consists of three equations. 

\begin{figure}[t]
	\begin{center}
    \subfigure{	
		  \input{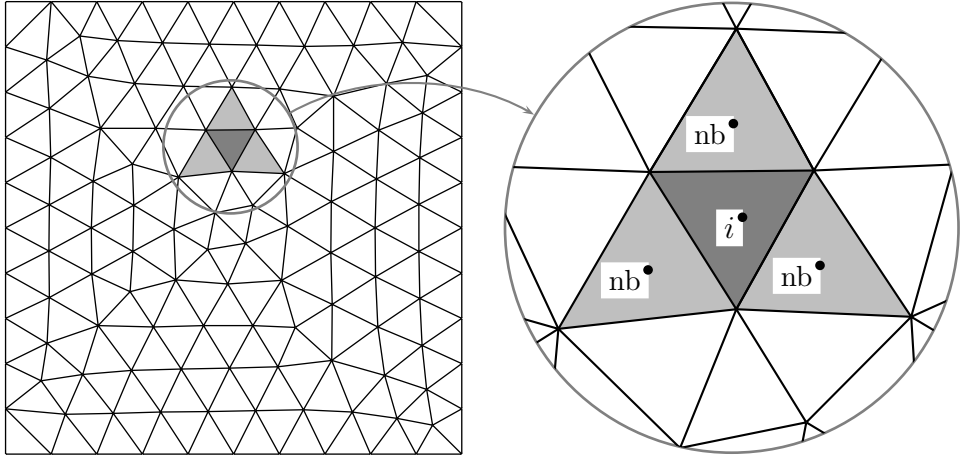}
		}
		\subfigure{	
		  \psset{unit=20.0cm}
\begin{pspicture}(0.343,0.529)(0.643,0.829)
	\psclip{\pscircle[linecolor=gray,linewidth=1pt](0.493,0.679){0.15}}
	\pspolygon[linestyle=none,fillstyle=solid,fillcolor=lightgray]%
		(0.379,0.612)(0.496,0.625)(0.439,0.716)
	\pspolygon[linestyle=none,fillstyle=solid,fillcolor=gray]%
		(0.496,0.625)(0.439,0.716)(0.547,0.717)
	\pspolygon[linestyle=none,fillstyle=solid,fillcolor=lightgray]%
		(0.439,0.716)(0.547,0.717)(0.495,0.811)
	\pspolygon[linestyle=none,fillstyle=solid,fillcolor=lightgray]%
		(0.496,0.625)(0.547,0.717)(0.611,0.620)
	\psset{linewidth=0.5pt}
	\psset{linewidth=0.8pt}
       \psline{cc-cc}(0.278,0.641)(0.379,0.612)
       \psline{cc-cc}(0.329,0.720)(0.439,0.716) 
       \psline{cc-cc}(0.369,0.497)(0.379,0.612)
       \psline{cc-cc}(0.379,0.612)(0.496,0.625)
       \psline{cc-cc}(0.379,0.612)(0.329,0.720)
       \psline{cc-cc}(0.379,0.612)(0.281,0.550)
       \psline{cc-cc}(0.439,0.716)(0.379,0.612)
       \psline{cc-cc}(0.439,0.716)(0.496,0.811) 
       \psline{cc-cc}(0.439,0.716)(0.390,0.813) 
       \psline{cc-cc}(0.459,0.533)(0.379,0.612)
       \psline{cc-cc}(0.496,0.625)(0.611,0.620)
       \psline{cc-cc}(0.496,0.625)(0.459,0.533) 
       \psline{cc-cc}(0.496,0.625)(0.439,0.716) 
       \psline{cc-cc}(0.496,0.811)(0.547,0.717)        
       \psline{cc-cc}(0.542,0.550)(0.496,0.625)
       \psline{cc-cc}(0.547,0.717)(0.496,0.625) 
       \psline{cc-cc}(0.547,0.717)(0.439,0.716)  
       \psline{cc-cc}(0.547,0.717)(0.601,0.807) 
       \psline{cc-cc}(0.547,0.717)(0.639,0.721)
       \psline{cc-cc}(0.547,0.717)(0.496,0.625) 
       \psline{cc-cc}(0.611,0.620)(0.547,0.717)  
       \psline{cc-cc}(0.611,0.620)(0.618,0.504) 
       \psline{cc-cc}(0.611,0.620)(0.542,0.550) 
       \psline{cc-cc}(0.611,0.620)(0.719,0.669)        
       \psline{cc-cc}(0.639,0.721)(0.611,0.620) 
       \psline{cc-cc}(0.714,0.558)(0.611,0.620) 
       \psline{cc-cc}(0.496,0.811)(0.550,0.904) 
       \psline{cc-cc}(0.447,0.906)(0.496,0.811) 
       \psline{cc-cc}(0.439,0.716)(0.496,0.811) 
       \psline{cc-cc}(0.496,0.811)(0.547,0.717) 
       \psline{cc-cc}(0.496,0.811)(0.390,0.813) 
       \psline{cc-cc}(0.601,0.807)(0.496,0.811)
       \psline{cc-cc}(0.526,0.466)(0.542,0.550)
       \psline{cc-cc}(0.542,0.550)(0.618,0.504)
       \psline{cc-cc}(0.459,0.533)(0.542,0.550) 
	\rput*[rt](0.433,0.651){\small $\rm nb$}
	\rput*[rt](0.489,0.748){\small $\rm nb$}
	\rput*[rt](0.546,0.654){\small $\rm nb$}
	\rput*[rt](0.495,0.686){$i$}
	\psdots(0.438,0.651)(0.494,0.748)(0.551,0.654)(0.500,0.686)
	\endpsclip
	\pnode(0.363,0.754){B}
\end{pspicture}
		}
	  \nccurve[angleA=30,angleB=150,linecolor=gray]{->}{A}{B}				
	\end{center}
	\caption{Schematic of a control volume and its neighbors in the 
	cell-centered finite volume method on an unstructured triangular mesh}
	\label{fig:1}
\end{figure}

The first step in a finite volume solution is to discretize the governing equations on a mesh. 
In this work, we use the cell-centered finite volume method on unstructured meshes with arbitrarily 
shaped cells, shown in Figure~\ref{fig:1}. The discretization starts with integrating 
Equations~\eqref{eq:1} and~\eqref{eq:2} over control volumes to obtain the integral form of the 
governing equations. For this purpose we use a cell-centered finite volume approach, shown in 
Figure~\ref{fig:1}. Then we use the Gauss theorem to transform the volume integrals into surface 
integrals.
\begin{gather}
\oint_{\partial {\Omega_i}} \rho {\bf v} \cdot \hat{\bf n}\, dA = 0 \label{eq:3} \\
\oint_{\partial {\Omega_i}} ( \rho {\bf v}\otimes{\bf v} ) \cdot \hat{\bf n}\, dA =
-\oint_{\partial {\Omega_i}} p \, \hat{\bf n}\, dA + 
\oint_{\partial {\Omega_i}} \mu(\nabla {\bf v} + \nabla {\bf v}^T ) \cdot \hat{\bf n}\, dA \label{eq:4} 
\end{gather}
where ${\Omega_i}$ is the $i$-th control volume and $\partial {\Omega_i}$ is its boundary, or rather its faces.

The next step in the discretization process is to approximate the surface integrals of Equations~\eqref{eq:5} 
and~\eqref{eq:6} in terms of the average flow across the control volumes faces. 
\begin{gather}
\sum_{\rm faces} J_{\rm mass} = 0 \label{eq:5} \\
\sum_{\rm faces} {\bf J}_{\rm adv} = 
-\sum_{\rm faces} {\bf F}_{\rm pres} + \sum_{\rm faces} {\bf F}_{\rm visc} \label{eq:6}
\end{gather}
where $J_{\rm mass}$ is the mass flow rate, 
${\bf J}_{\rm mom}$ is the vector of momentum flow rate, 
and ${\bf F}_{\rm pres}$ and ${\bf F}_{\rm visc}$ are the vectors of pressure and
viscous forces, respectively. Note that the values of these flows and forces at each face are unknown and
need to be determined.

\begin{figure}
	\begin{center}
	  \includegraphics[width=.5\textwidth]{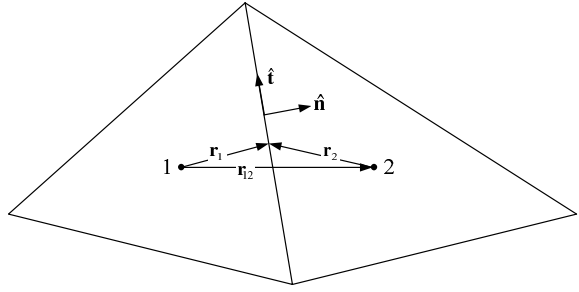}
	\end{center}
	\caption{Schematic of an interior face and its neighboring control volumes}
	\label{fig:2}
\end{figure}

To determine the values of mass and momentum flows across a face and those of forces acting on it,
we need to interpolate the pressure and velocities from the nodes of the neighboring control volumes.
For example if we use a second-order central discretization scheme, the values of a typical variable, $\psi$,
and its gradient, $\nabla\psi$, at the interface of control volumes 1 and 2, shown in Figure~\ref{fig:2}, 
become~\cite{baserinia:2008}:
\begin{gather}
  \psi_f \approx \frac{1}{2}(\psi_1+\psi_2) + \frac{1}{4} ( \nabla \psi_1 + 
  \nabla \psi_2 ) \cdot ( {\bf r}_1 + {\bf r}_2 ) + {\cal O}(h^2) \label{eq:7} \\
  \nabla \psi_f \approx \frac{1}{2} ( \nabla \psi_1 + \nabla \psi_2 ) + {\cal O}(h) \label{eq:8}
\end{gather}
where $h$ is the local mesh characteristic size. Following Ferziger and Peric~\cite{ferziger:2002} 
and Zwart~\etal~\cite{zwart:1999}, we may discretize the face-normal component of the gradient 
vector differently to improve the numerical stability of the scheme:
\begin{equation}
  \left. \frac{\partial \psi}{\partial n}\right|_f \approx (\nabla \psi \cdot \hat{\bf n})_f \approx 
  \alpha\left(\frac{\psi_2-\psi_1}{|{\bf s}|} \right) + 
  \nabla \psi_f \cdot (\hat{\bf n}-\alpha \hat{\bf s}) + {\cal O}(h)\label{eq:9}
\end{equation}
where $\alpha = \hat{\bf n} \cdot \hat{\bf s}$ represents the mesh nonorthogonality~\cite{zwart:1999}.
Note that the approximations in Equations~\eqref{eq:8} and~\eqref{eq:9} are formally first-order
accurate on a nonuniform mesh. However both approximations exhibit a second-order convergence as the
mesh is refined~\cite{ferziger:2002}. Therefore the overall order of accuracy of the discretization
method would still be two.

Using Equations~(\ref{eq:7}--\ref{eq:9}), we can approximate the values
of velocity and pressure at an interior face. For a boundary face we may use a similar approach but
the central discretization scheme must be replaced by a one-sided extrapolation scheme~\cite{baserinia:2008}.
Once the velocity and pressure values are determined at a face, we can calculate the mass and momentum
flows and pressure and viscous forces at the face.

To calculate the mass flow rate across a face, we use the Rhie and Chow velocity pressure interpolation
scheme~\cite{rhie:1983}. This scheme prevents the infamous velocity-pressure decoupling from happening.
The result is:
\begin{equation}
  J_{{\rm mass}_{12}} = \int_{\rm face} \rho {\bf v} \cdot \hat{\bf n}\, dA \approx \rho\left[ \bar{V}_n - d_f 
    \left( \left. \frac{\partial p}{\partial n} \right|^{\rm active}_{\rm face} -
           \left. \frac{\partial p}{\partial n} \right|^{\rm lagged}_{\rm face} 
    \right)\right]
  \label{eq:10}
\end{equation}
where $\bar{V}_n$ is the advected velocity and $d_f$ is the dissipation coefficient~\cite{reuss:2002,zwart:1999}.
Using Equations~\eqref{eq:7}, \eqref{eq:8}, and~\eqref{eq:9} for calculating $\bar{V}_n$,
${\partial p}/{\partial n}|^{\rm lagged}$, and ${\partial p}/{\partial n}|^{\rm active}$, respectively,
results in the discretized form of the mass flow across an interior face:
\begin{equation}
  \begin{split}
  J_{{\rm mass}_{12}} =\;& \frac{\rho A_f}{2} ({\bf v}_1 + {\bf v}_2) \cdot \hat{\bf n} +
  \frac{\rho \alpha d_f A_f}{|\bf{s}|} ( p_1 - p_2) + \\
  & \frac{\rho A_f}{4} [ \nabla ({\bf v}_1\cdot\hat{\bf n})  + \nabla ({\bf v}_2\cdot\hat{\bf n})] \cdot
   ({\bf r}_1 + {\bf r}_2) +
  \frac{\rho \alpha d_f A_f}{2}  (\nabla p_1 + \nabla p_2 ) \cdot \hat{\bf s}
  \end{split} 
  \label{eq:11}
\end{equation}
where $A_f$ is the face area in Fiqure~\ref{fig:2}.
To calculate the momentum flow vector across a face, ${\bf J}_{\rm adv}$, we need to use
the upwind approximation:
\begin{equation}
  {\bf J}_{{\rm adv}_{12}} = \int_{\rm face} ( \rho {\bf v}\otimes{\bf v} ) \cdot \hat{\bf n}\, dA \approx
    \rho J_{{\rm mass}_{12}} ({\bf v}_{\rm up} + \nabla {\bf v}_{\rm up} \cdot {\bf r}_{\rm up})
\end{equation}
where ${\bf v}_{\rm up}={\bf v}_{1}$ if $J_{{\rm mass}_{12}}>0$ and ${\bf v}_{\rm up}={\bf v}_{2}$
otherwise. We can also calculate the pressure and viscous forces acting on a face using 
Equations~(\ref{eq:7}--\ref{eq:9}):
\begin{align}
  {\bf F}_{{\rm pres}_{12}} &= \int_{\rm face} p \, \hat{\bf n}\, dA \approx
  \left[ \frac{1}{2} (p_1+p_2) + \frac{1}{4}(\nabla p_1 + \nabla p_2) \cdot ({\bf r}_1+{\bf r}_2) \right]\hat{\bf n}A_f \\
  {\bf F}_{{\rm visc}_{12}} &= \int_{\rm face} \mu(\nabla {\bf v} + \nabla {\bf v}^T ) \cdot \hat{\bf n}\, dA \nonumber \\
  &\approx \mu\left[\alpha\left(\frac{{\bf v}_2-{\bf v}_1}{|{\bf s}|} \right) + 
  \frac{1}{2}(\nabla {\bf v}_1+\nabla{\bf v}_2) \cdot (\hat{\bf n}-\alpha \hat{\bf s}) +
  \frac{1}{2}(\nabla {\bf v}_1^T+\nabla{\bf v}_2^T) \cdot \hat{\bf n}
  \right] A_f \label{eq:14}
\end{align}
Substituting Equations~(\ref{eq:11}--\ref{eq:14}) into Equations~\eqref{eq:5} and~\eqref{eq:6} 
results in the discretized form of the mass and momentum equations. Although the final result
is tedious, we can represent it in the following symbolic form:
\begin{eqnarray}
  \phi_i + \sum_{j \in {\rm nb}} c_{ij} \phi_j = b_i
  \label{eq:15}
\end{eqnarray}
where $\phi_i = (p_i, u_i, v_i)^T$ is the vector of primitive variables at the $i$-th control volume 
and $j$ is the index of its neighboring control volumes, denoted by `nb' in Figure~\ref{fig:1}.
Note that in Equation~\eqref{eq:15}, the coefficients~$c_{ij}$ and~$b_i$
are generally functions of $\phi$. Therefore it is a nonlinear equation which should be solved using 
iterative methods.

\section{Truncation Error in the Finite Volume Method}
Let us describe the concept of truncation error in the finite volume
context. Suppose we represent the integral form of the governing equations, 
Equations~\eqref{eq:3} and~\eqref{eq:4}, in the following symbolic form:
\begin{equation}
{\mathscr L}_h(\Phi) = 0
\label{eq:16}
\end{equation}
where ${\mathscr L}_h$ is the integral conservation operator on a mesh with the characterisitic size $h$,
acting on the exact solution vector, $\Phi=(p,u,v)^{\rm T}$. 
Nevertheless in the finite volume method we solve the discretized form of the above equation. In the case
of a second-order discretization, which is the most common choice in CFD applications, the symbolic
form of the discretized equation is:
\begin{equation}
  {\cal L}_h^2(\phi_h^2) = 0
\end{equation}
where ${\cal L}_h^2$ is the second-order discrete conservation operator on a mesh with the 
characteristic size $h$ and $\phi_h^2$ is numerical solution. The difference between the 
numerical solution, $\phi_h^2$, and the exact solution, $\Phi$, is the discretization error. 
\begin{equation}
\varepsilon_h^2 = \phi_h^2 - \Phi
\end{equation}
Note that the numerical solution, $\phi_h^2$, may not satisfy the exact conservation 
operator, ${\mathscr L}_h$. Therefore substituting the numerical solution, $\phi_h^2$, into 
Equation~\eqref{eq:16} results in an apparent source term, called the truncation error and shown by $\delta_h^2$:
\begin{equation}
{\mathscr L}_h(\phi_h^2) = \delta_h^2
\label{eq:19}
\end{equation}
The latter shows that the truncation error of a numerical solution is the
apparent source term required in the exact equations to satisfy the numerical solution. 
On the other hand, using the divergence theorem on a control volume, we can prove that 
the apparent source term is equal to the net flow imbalance across the boundaries of the 
control volume. Therefore the truncation error is simply the net flow error across the boundaries of a 
control volume. This definition of truncation error is on a volume-based basis, contrary to the 
face-based definitions in Roe and Nishikawa~\cite{roe:2002} and Reuss and Stubley~\cite{reuss:2003}. 

\section{Calculating the Volume-Based Truncation Error}
In the finite volume context, the truncation error is equal to the net imbalance of mass 
and momentum flows across the faces of each control volume as mentioned in the previous section.
Although this definition of truncation error is based on Equation~\eqref{eq:19}, we cannot simply
use it to calculate the volume-based truncation error of the numerical solution, $\phi_h^2$. Note
that $\phi_h^2$ is only peace-wise continuous within control volumes but discontinuous 
across control volume faces. Therefore it is not feasible to evaluate ${\mathscr L}_h(\phi_h^2)$ 
since it involves evaluating integrals on the faces.

\begin{figure}
	\begin{center}
    \subfigure[Exact flow]{	
      \label{fig:3-a}
		  \includegraphics[width=0.22\textwidth]{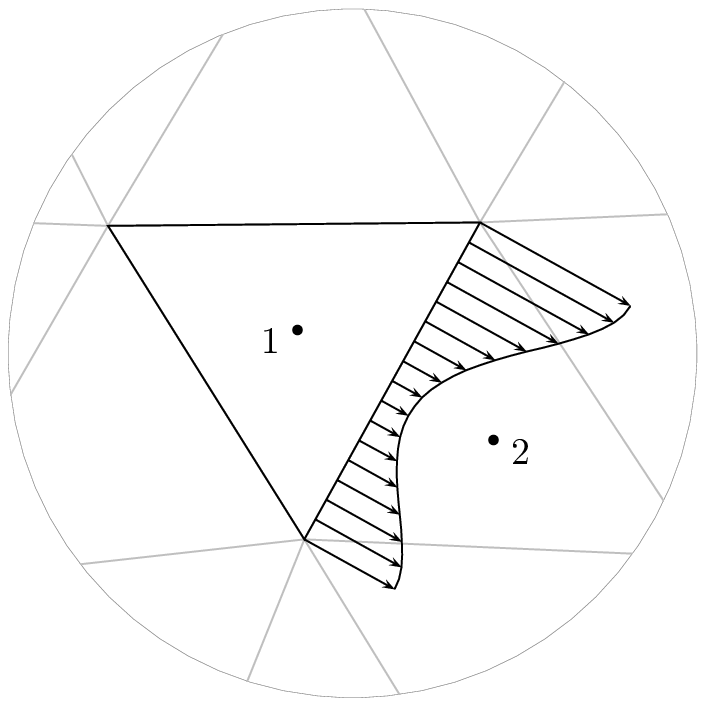}
		}
		\hfill
		\subfigure[Approximate flow, 2nd order method]{	
  		\label{fig:3-b}
		  \includegraphics[width=0.22\textwidth]{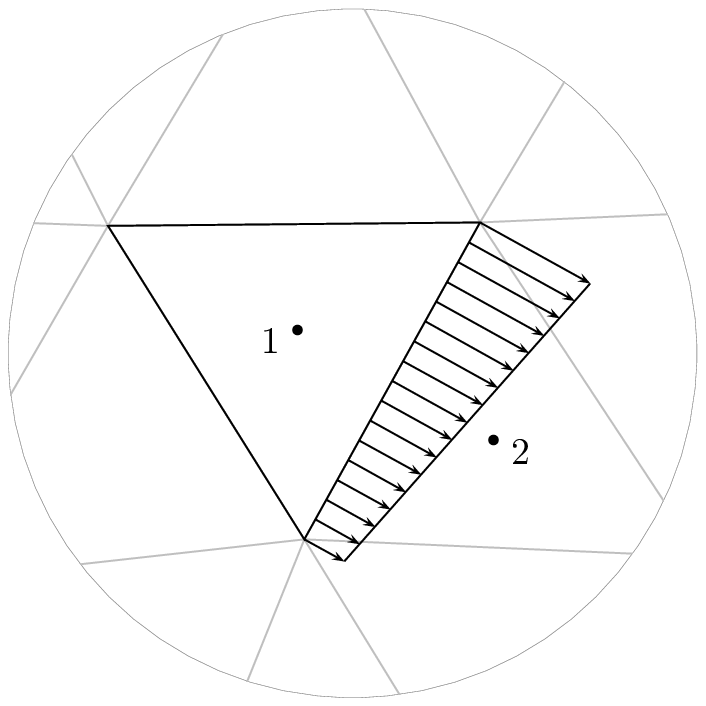}
		}
		\hfill
    \subfigure[Approximate flow, 2nd order method on a refined mesh]{	
      \label{fig:3-c}
		  \includegraphics[width=0.22\textwidth]{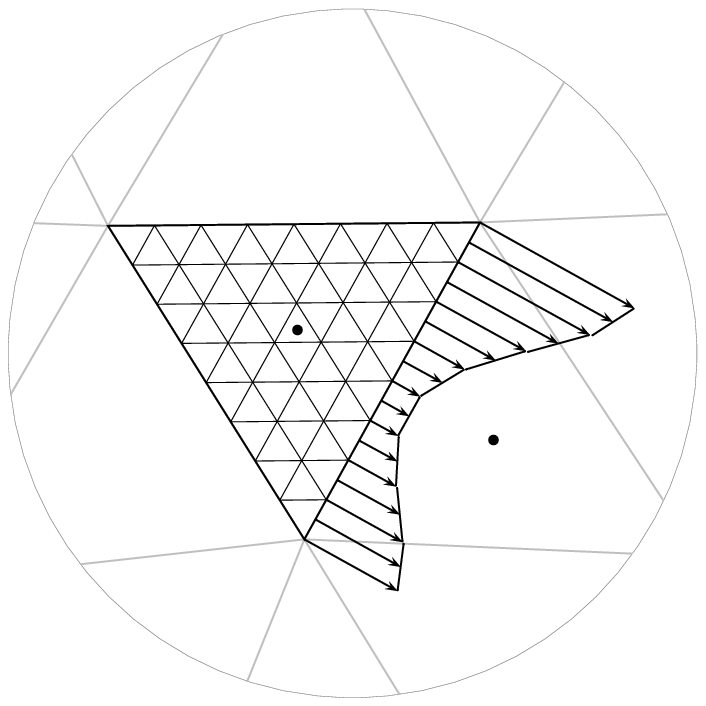}
		}
		\hfill
    \subfigure[Approximate flow, higher order method]{	
      \label{fig:3-d}
		  \includegraphics[width=0.22\textwidth]{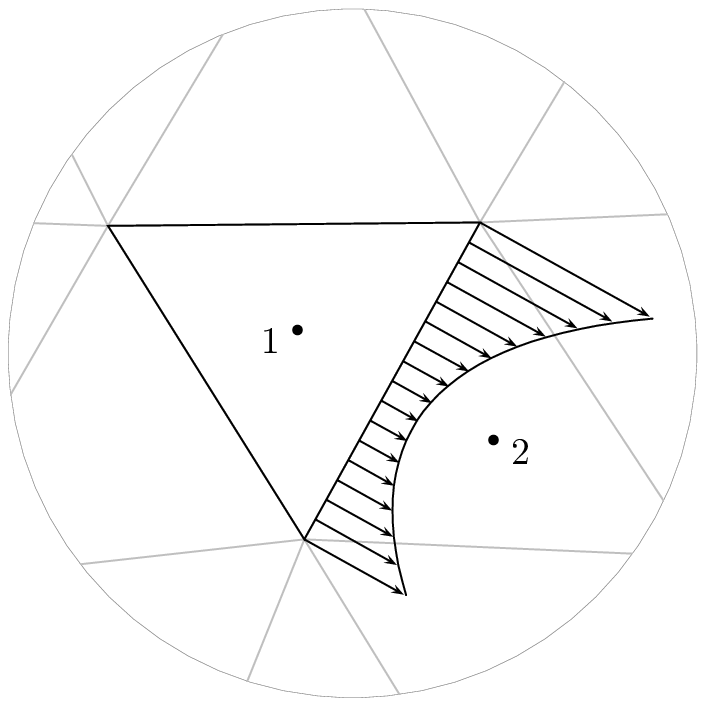}
		}
	\end{center}
	\caption{Comparison of exact face flow and its approximation based on
	various numerical schemes and mesh sizes}
	\label{fig:3}
\end{figure}

The alternative approach to truncation error calculation is to replace the exact operator ${\mathscr L}_h$ by 
an accurate discrete operator. Due to the consistency condition~\cite{chung:2002}, the most obvious choice 
for an accurate discrete operator is ${\cal L}_{h^\prime}^{2}$ where $h^\prime \ll h$; in
other words the second-order discrete operator on a very fine mesh. To 
understand how to estimate the truncation error using this operator, imagine each face of a 
two-dimensional mesh is divided into $n$ smaller faces. Therefore each control volume is split
into $n^2$ smaller control volumes, shown in Figures~\ref{fig:3-b} and~\ref{fig:3-c}. If we interpolate
the second-order solution $\phi_h^2$ from the original mesh of Figure~\ref{fig:3-b} to the fine mesh
of Figure~\ref{fig:3-c} and apply the conservation operator ${\cal L}_{h^\prime}^{2}$ on it, we can
obtain the truncation errors of individual control volumes on the fine mesh. To obtain the truncation error
on the original mesh of Figure~\ref{fig:3-b}, it suffices to find the summation of truncation errors of
the smaller control volumes in Figure~\ref{fig:3-c}. The reason is that the face flow errors
of the smaller control volumes cancel out each other except at the faces of the original control
volume of Figure~\ref{fig:3-b}, which provides the truncation error. Unfortunately this approach is both
hard to implement and intense on computer memory requirement. Therefore it is rarely used in
practical applications.

In the literature, there are three techniques for estimating the face flow errors~\cite{hay:2006}:
\begin{itemize}
\item Estimating the neglected terms in the  discretization scheme;
\item Recovering a higher order accurate solution based on the discrete solution; and
\item Use of a higher order accurate discretization scheme.
\end{itemize}

The first technique for estimating the face flow errors  is based on analyzing 
the neglected terms in the Taylor series expansion in the discretization scheme.
The neglected terms are estimated and used for establishing a
higher order accurate face flow calculation. Therefore these neglected terms
provide an estimate for the mass and momentum face flow errors. Examples of
this technique in the literature are the works of Ilinca \etal~\cite{ilinca:2000},
Reuss and Stubley~\cite{reuss:2003}, and Zhang \etal~\cite{zhang:2000}. 

The second technique for estimating the face flow errors is based on recovering a 
higher order accurate solution from the available discrete solution in order
to obtain a more accurate face flow estimation. In this technique, the discrete 
solution on an initial mesh is used along with a higher order accurate 
interpolation scheme. Since the discrete solution is only given at the control
volume nodes, the higher order interpolation provides a more accurate estimate
of the solution variables at the faces. Therefore an estimate for the
face flow error can be obtained. Examples of this method can be seen in
recent works of Hay and Visonneau~\cite{hay:2006} and Hay \etal~\cite{hay:2006-2}

The third technique for estimating the face flow errors is based on using a higher
order discretization scheme. In this technique, one uses the numerical solution
obtained from an $n$-th order accurate discretization scheme along with
another discretization scheme which is $m$-th order accurate where $m>n$.
Substitution of the $n$-th order solution into the $m$-th order discretization
scheme results in an estimate of the truncation error. This technique, which is 
referred to as the defect correction method, is elaborated on by Ervin and 
Layton~\cite{ervin:1989} and Pierce and Giles~\cite{pierce:2004}.

Note that all the above techniques are conceptually equivalent. Figure~\ref{fig:3-d}
shows the basic concept of these techniques. As seen, these techniques
try to estimate the face flow errors of Figure~\ref{fig:3-b} by using higher order
terms in the discretization scheme. The advantage of
the first two techniques is the ease of implementation while the third technique
is more cumbersome to develop and implement. In addition, the first technique
does not involve complicated interpolations, which is an advantage in terms of
implementation especially on unstructured meshes. Therefore we use the first 
technique to estimate the face flow errors.

\section{The Proposed Truncation Error Estimation Scheme}
This section explains a truncation error estimation scheme that is based on evaluating 
the first neglected Taylor series terms in the calculation of the mass and momentum
flow errors. A second-order accurate discretization scheme uses Equations~(\ref{eq:7}--\ref{eq:9})
to calculate the face flows. Therefore one needs to account for the first neglected terms
in these equations: 
\begin{gather}
  \psi_f \approx \frac{1}{2}(\psi_1+\psi_2) + \frac{1}{4} ( \nabla \psi_1 + 
  \nabla \psi_2 ) \cdot ( {\bf r}_1 + {\bf r}_2 ) + \nonumber \\
  \frac{1}{4} {\bf r}_1^T \cdot (\nabla \nabla \psi_1+\nabla \nabla \psi_2) \cdot {\bf r}_2 + {\cal O}(h^3) \label{eq:20}\\
  \nabla \psi_f \approx \frac{1}{2} ( \nabla \psi_1 + \nabla \psi_2 ) +
  \frac{1}{4} ( \nabla\nabla \psi_1 +   \nabla\nabla \psi_2 ) \cdot ( {\bf r}_1 + {\bf r}_2 )  + {\cal O}(h^2) 
\end{gather}
\begin{gather}
  \left. \frac{\partial \psi}{\partial n}\right|_f = (\nabla \psi \cdot \hat{\bf n})_f \approx 
  \alpha\left(\frac{\psi_2-\psi_1}{|{\bf s}|} \right) + 
  \nabla \psi_f \cdot (\hat{\bf n}-\alpha \hat{\bf s}) + \nonumber\\
  \frac{1}{4}({\bf r}_1 + {\bf r}_2)^T \cdot (\nabla\nabla\psi_1+\nabla\nabla\psi_2) \cdot \hat{\bf n} + {\cal O}(h^2)\label{eq:22}
\end{gather}
In these equations $\nabla\nabla \psi$ is the tensor of second-order derivatives, called the Hessian tensor.
Using the equations above and following the procedure of Section~3, one can derive new equations
for the mass and momentum flows across a face. These relations would be similar to 
Equations~(\ref{eq:10}--\ref{eq:14}) with some extra terms, which account for the face flow errors.
The final result for the mass face flow error across an interior face, shown in Figure~\ref{fig:2}, is:
\begin{equation}
  \Delta J_{\rm mass} \approx \frac{\rho A_f}{4} \left[ \nabla \nabla ({\bf v}_1\cdot\hat{\bf n}) + 
                            \nabla \nabla ({\bf v}_2\cdot\hat{\bf n}) \right]: 
        ( {\bf r}_1 \otimes {\bf r}_2 + \frac{1}{12} {\bf t} \otimes {\bf t} ) \label{eq:23}
\end{equation}
where ${\bf t} = A_f \hat{\bf t}$ in Figure~\ref{fig:2}, and the symbols $(\otimes)$ and $(:)$ represent the tensor product
of two vectors and the Frobenius product of two tensors, respectively. Similarly one can obtain
the face flow and force errors in the momentum equation:
\begin{align}
  \Delta {\bf J}_{\rm adv} \approx\;& \frac{\rho A_f {\bf v}_{\rm up}}{4} \left[ \nabla \nabla ({\bf v}_1\cdot\hat{\bf n}) + 
                            \nabla \nabla ({\bf v}_2\cdot\hat{\bf n}) \right]: 
        ( {\bf r}_1 \otimes {\bf r}_2 + \frac{1}{12} {\bf t} \otimes {\bf t} ) + \nonumber \\
   & \frac{\rho A_f}{4} ({\bf v}_1 \cdot \hat{\bf n} + {\bf v}_2 \cdot \hat{\bf n}) 
  \left[\nabla \nabla {\bf v}_{\rm up} : 
    \left( {\bf r}_{\rm up} \otimes {\bf r}_{\rm up} + 
    \frac{1}{12}{\bf t} \otimes {\bf t} \right)\right] + \nonumber \\
  & \frac{\rho A_f}{24} \left[ \nabla ({\bf v}_1\cdot\hat{\bf n}) + \nabla ({\bf v}_2\cdot\hat{\bf n})\cdot {\bf t} \right] 
  \cdot (\nabla {\bf v}_{\rm up} \cdot {\bf t}) + {\cal O}(h^3)\\
  \Delta {\bf F}_{\rm pres} \approx\;& \frac{\hat{\bf n} A_f }{4} ( \nabla \nabla p_1 + \nabla \nabla p_2 ) :
        \left( {\bf r}_1 \otimes {\bf r}_2 + \frac{1}{12} {\bf t} \otimes {\bf t} \right) + {\cal O}(h^3)\\
  \Delta {\bf F}_{\rm visc} \approx\;& \frac{\mu A_f}{4}({\bf r}_1 + {\bf r}_2)^T \cdot
  [ \nabla (\nabla {\bf v}_1 +\nabla {\bf v}_1^T) + 
  \nabla (\nabla {\bf v}_2 +\nabla {\bf v}_2^T) ]\cdot \hat{\bf n} + {\cal O}(h^2) \label{eq:26}
\end{align}
And the total face error for the momentum flow across a face based on 
the equation above and Equation~\eqref{eq:6} becomes:
\begin{equation}
  \Delta {\bf J}_{\rm mom} = \Delta {\bf J}_{\rm adv}+\Delta {\bf F}_{\rm pres}-\Delta {\bf F}_{\rm visc}\label{eq:27}
\end{equation}
Note that Equations~\eqref{eq:23} and~\eqref{eq:27} only provide the mass and momentum flow errors across
a face. To calculate the total truncation error for a control volume, one has to find the summation of 
flow errors across all the control volume faces.

Equations~\eqref{eq:23} and~\eqref{eq:27} provide a straightforward scheme for estimating
mass and momentum truncation errors in the context of finite volume methods on general unstructured meshes. 
The only problem with these equations is the requirement to calculate of the Hessian tensor of the 
primitive variables, which do not appear in a second-order discretization. However, this is not a
major issue since one can estimate the solution Hessians by applying a gradient reconstruction
procedure to the already known solution gradients~\cite{barth:1993, reuss:2003}.
In the following section, we apply the proposed truncation error scheme to two academic
test cases and discuss it performance.

\section{Results and Discussion}
To show the performance of the truncation error estimation scheme, discussed in the
previous section, we apply it to two academic test cases: incompressible laminar
flows in a lid-driven square cavity at ${\rm Re}=1000$ and over a backward facing step
at ${\rm Re}=400$. Then we discuss the strengths and shortcomings of the proposed scheme.

\subsection{Flow in a Lid-Driven Cavity}
The lid-driven cavity flow is an important test case since it exhibits many important
phenomena of fluid flows in a simple geometry. Figure~\ref{fig:3} shows the schematic 
of the lid-driven square cavity flow. The Reynolds number of the flow 
$\mbox{Re}= U L/\nu$, which is defined based on the lid velocity and the cavity size, is 1000.  
The boundary conditions of the flow are no-slip walls. To set the pressure level, the 
pressure at the lower left corner is set to zero. 

\begin{figure}
	\begin{center}
		\includegraphics[width=0.3\textwidth]{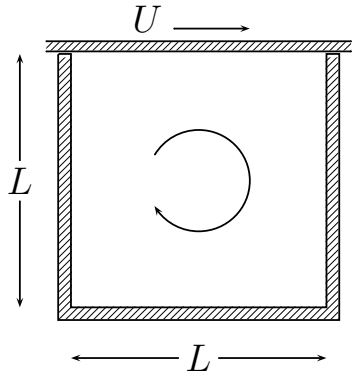}
	\end{center}
	\caption{Schematic of the lid-driven cavity flow at ${\rm Re}=\rho U L/\mu$}
	\label{fig:3}
\end{figure}

To obtain the basic second-order solution, we solve the cavity flow on a uniform unstructured 
triangular mesh with 3602 control volumes. To validate the truncation error estimates we have to find 
the actual volume-based truncation error
with a reasonable accuracy. For this purpose we interpolate the basic solution to a refined mesh
and then apply the second-order conservation operator on the refined mesh on the basic solution,
discussed in Section~3. Nevertheless we need to confirm that the truncation error obtained using this method
is accurate enough. For this purpose we use a mesh refinement
study by refining the mesh until the truncation error estimates converge. Based on this technique, if
we subdivide each face of the original mesh into 16 smaller faces, the RMS error in the truncation error
calculation would be smaller than one percent. Therefore we use the second-order conservation
operator on a mesh with 922112 control volumes as an approximation to the exact conservation 
operator.

\begin{figure}
  \begin{center}
    \subfigure[Actual mass truncation error]{
      \label{fig:5-a}
      \includegraphics[width=.45\textwidth]{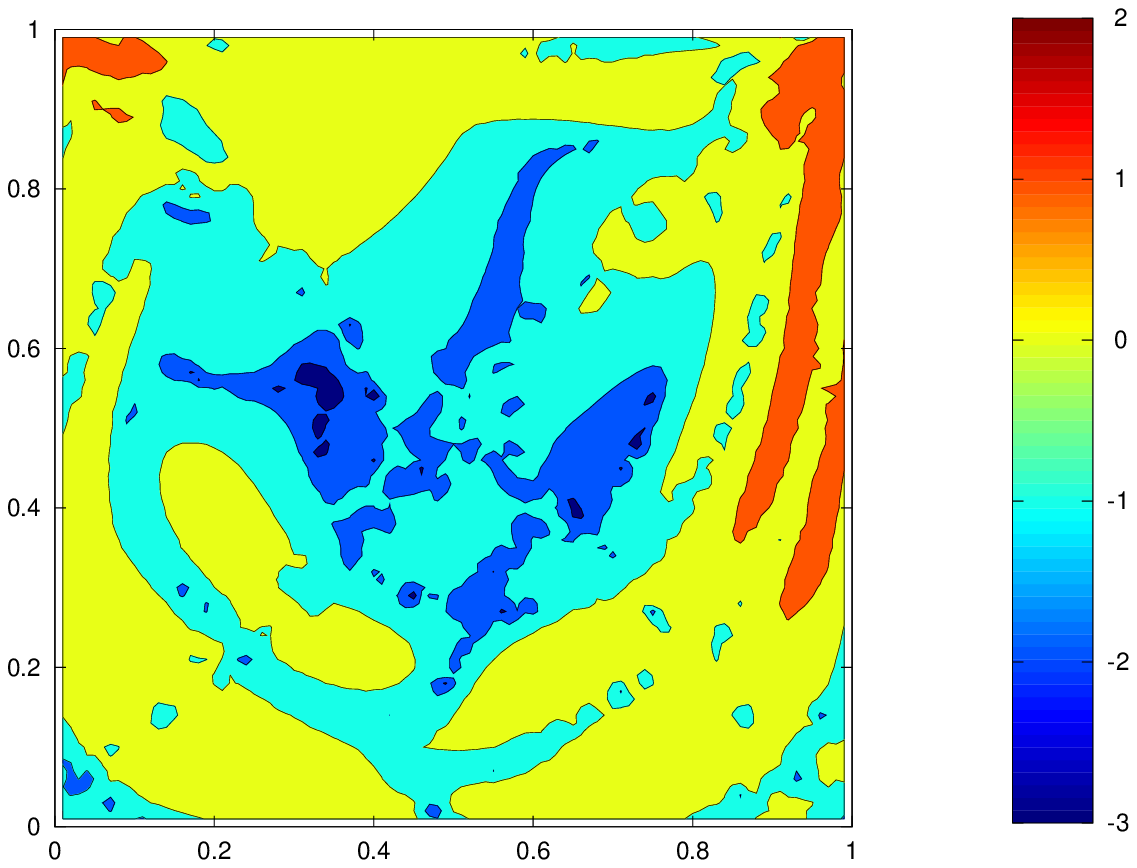}
    }
    \hfill
    \subfigure[Estimated mass truncation error]{
      \label{fig:5-b}
      \includegraphics[width=.45\textwidth]{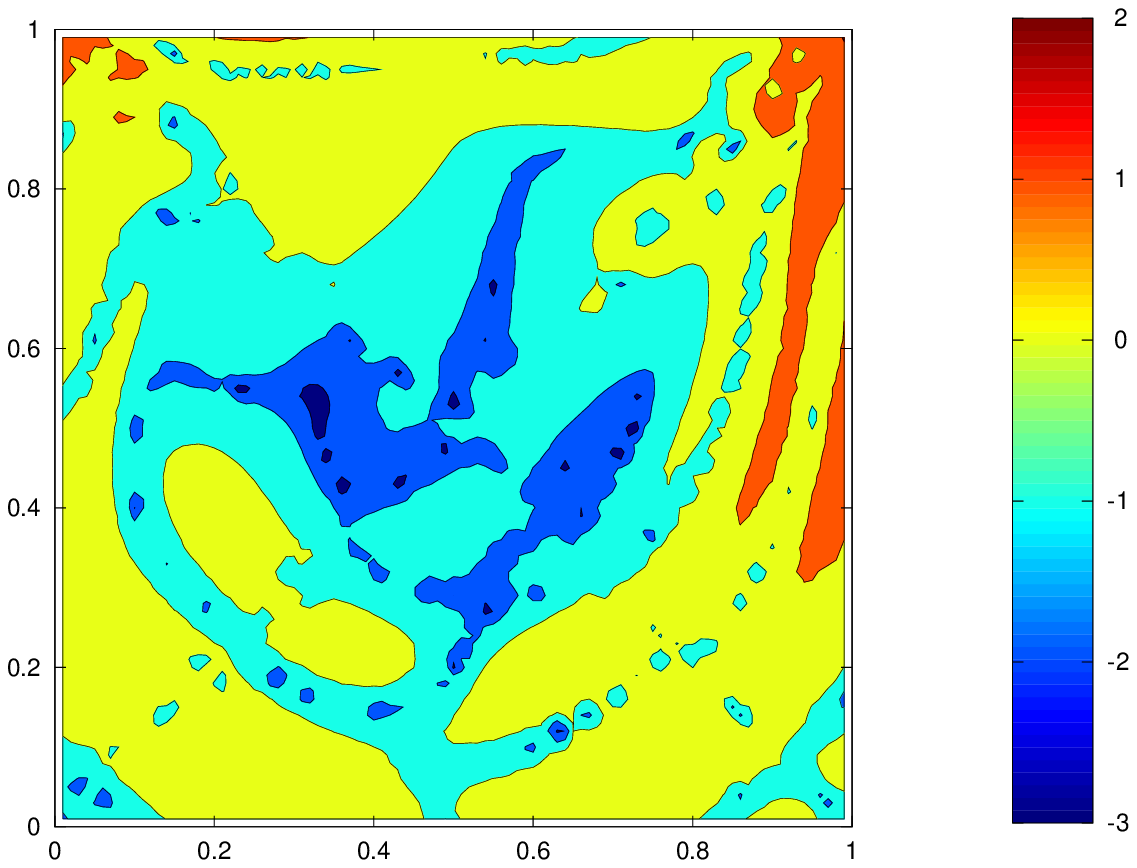}
    }
    \subfigure[Actual $x$-momentum truncation error]{
      \label{fig:5-c}
      \includegraphics[width=.45\textwidth]{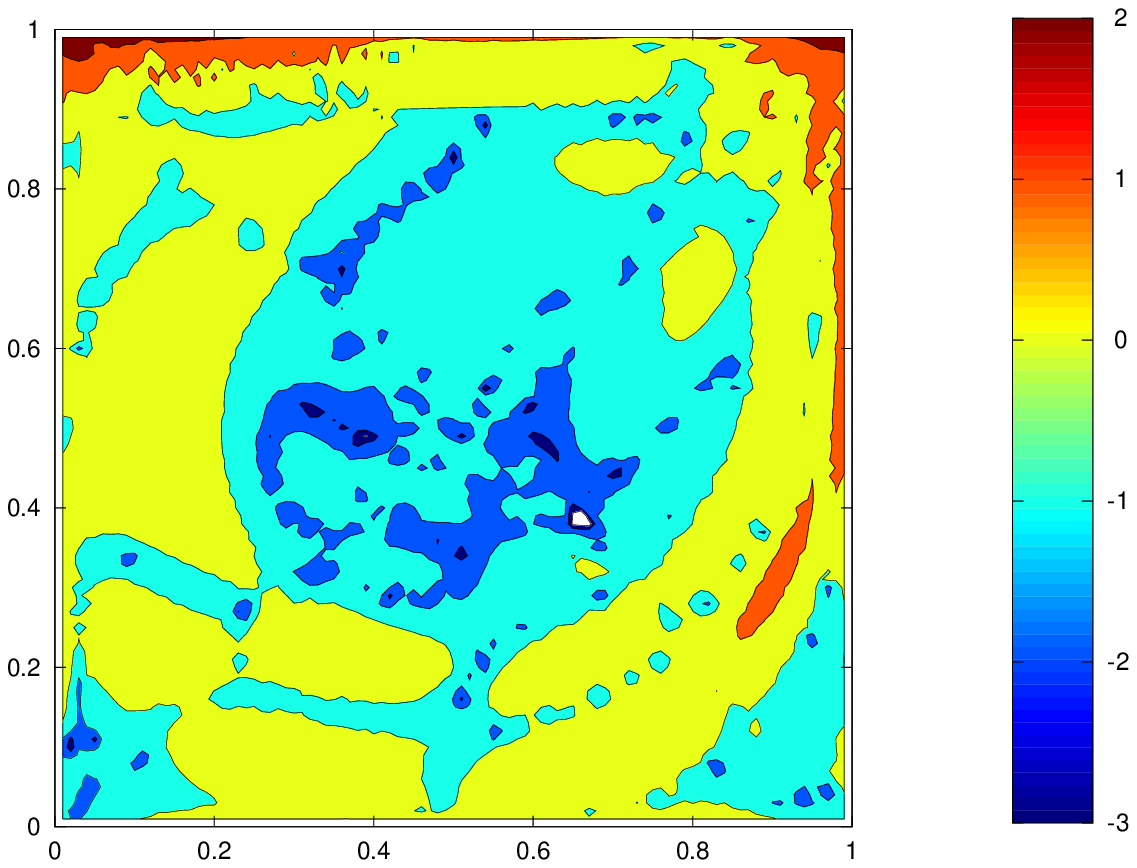}
    }
    \hfill
    \subfigure[Estimated $x$-momentum truncation error]{
      \label{fig:5-d}
      \includegraphics[width=.45\textwidth]{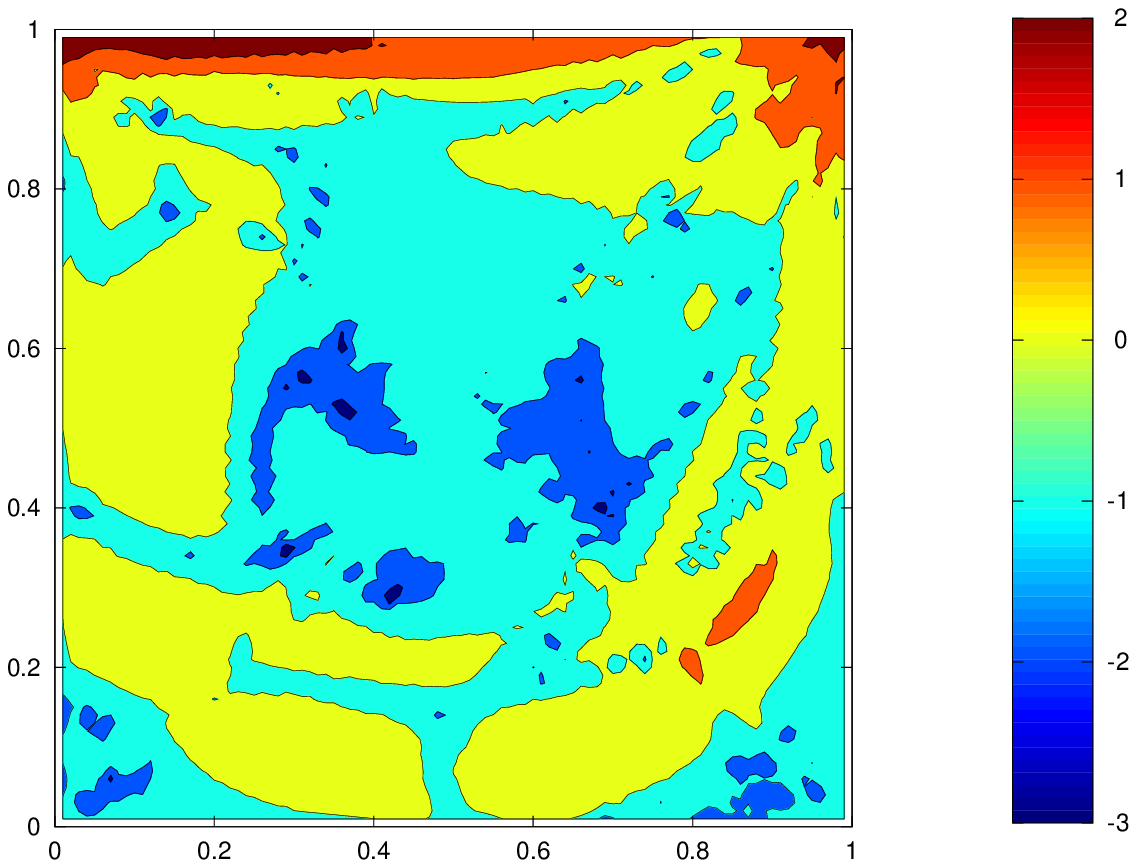}
    }
    \subfigure[Actual $y$-momentum truncation error]{
      \label{fig:5-e}
      \includegraphics[width=.45\textwidth]{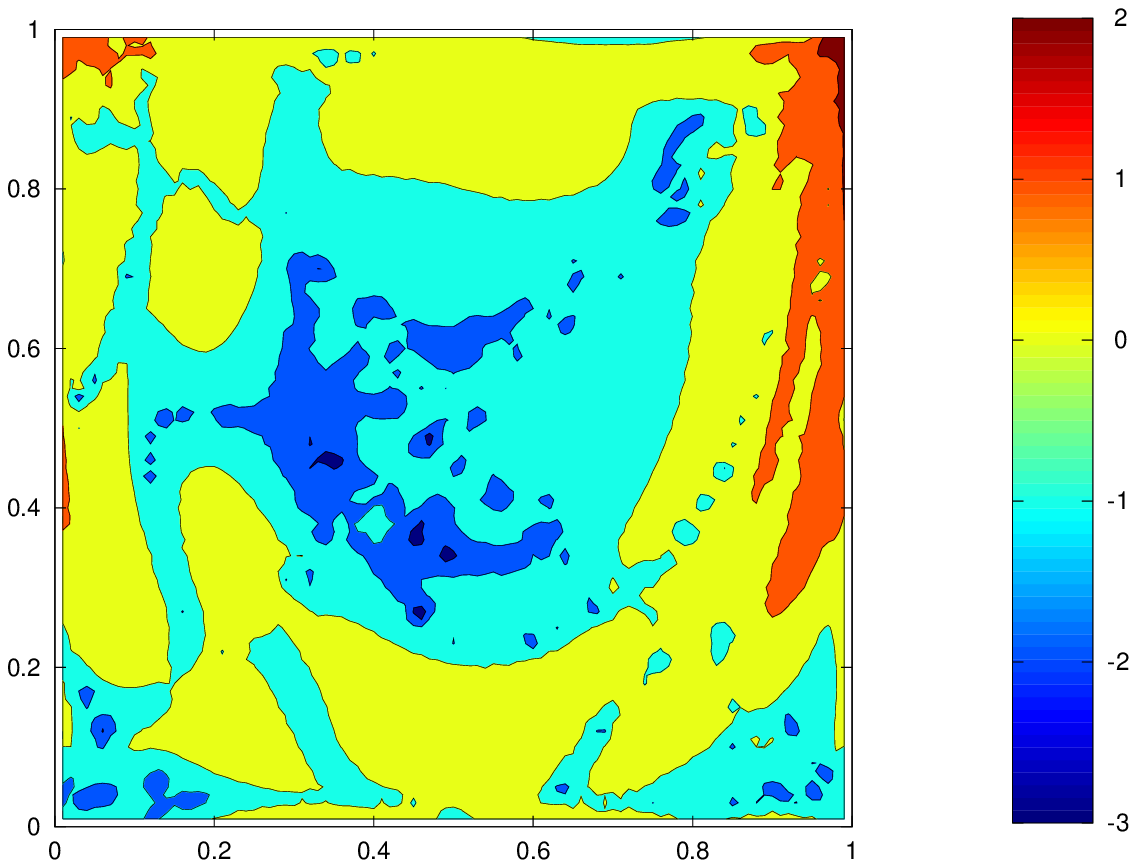}
    }
    \hfill
    \subfigure[Estimated $y$-momentum truncation error]{
      \label{fig:5-f}
      \includegraphics[width=.45\textwidth]{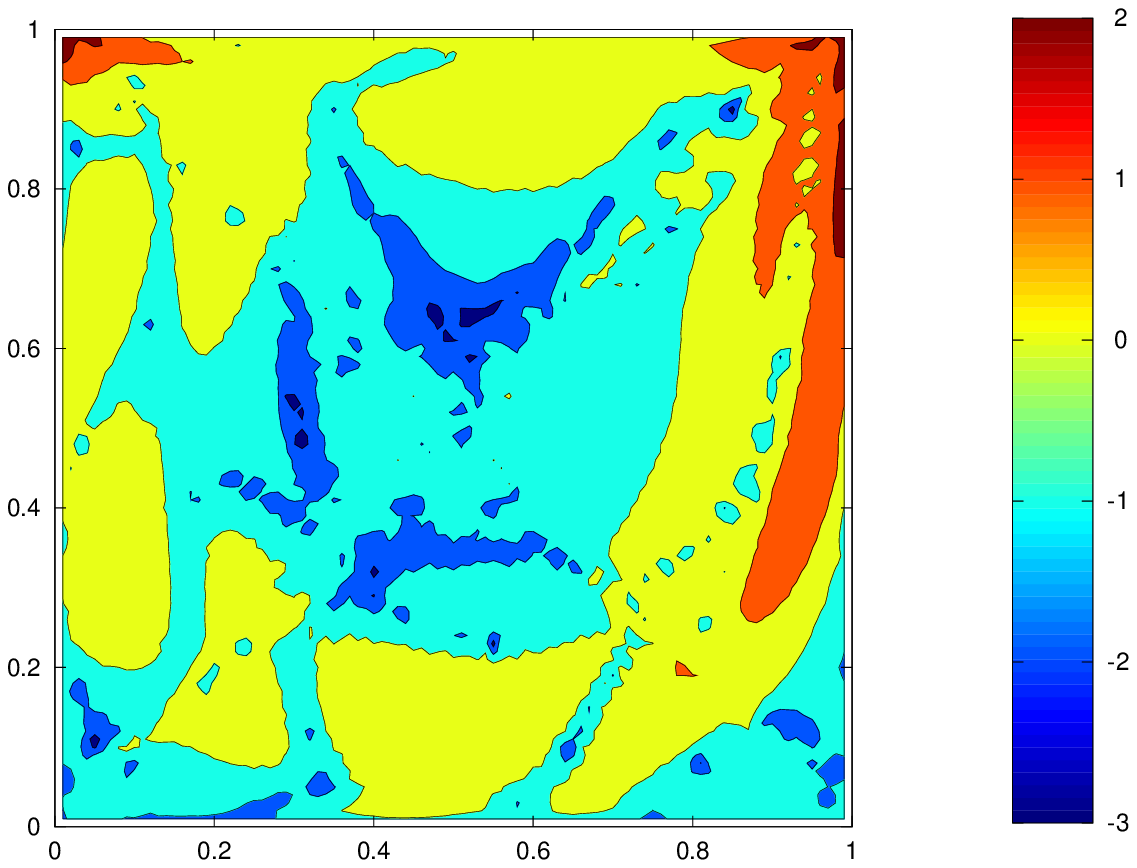}
    }
    \caption{Comparison between the actual and estimated truncation error distributions for
    the cavity flow at ${\rm Re}=1000$ on an isotropic triangular mesh.}
    \label{fig:5}
  \end{center}
\end{figure}

To obtain the truncation error estimates based on the proposed scheme we use 
Equations~\eqref{eq:23} and~\eqref{eq:27}. Figures~\ref{fig:5-a} through~\ref{fig:5-f}
show the distribution of the volume-based truncation error scaled by the average truncation error across
the domain. Note that the contour levels in these figure are logarithmic and as a result
these figures represent the truncation error order of magnitude.

A brief examination of Figure~\ref{fig:5} shows that the distributions of the estimated 
truncation errors are very similar to those of the actual truncation errors. In all cases, the truncation errors
near the driven lid and the right wall are orders of magnitude larger than those
of the rest of the domain. This is due to the strong velocity and pressure gradients
experienced in the near wall regions. In contrast, in the lower left and right corners
where two low energy separation bubbles form, the truncation errors are small. These observations
show that the proposed truncation error estimation scheme successfully predicts the overall trend
of the truncation error distribution.

\subsection{Flow over a Backward Facing Step}
\begin{figure}
	\includegraphics[width=1.0\textwidth]{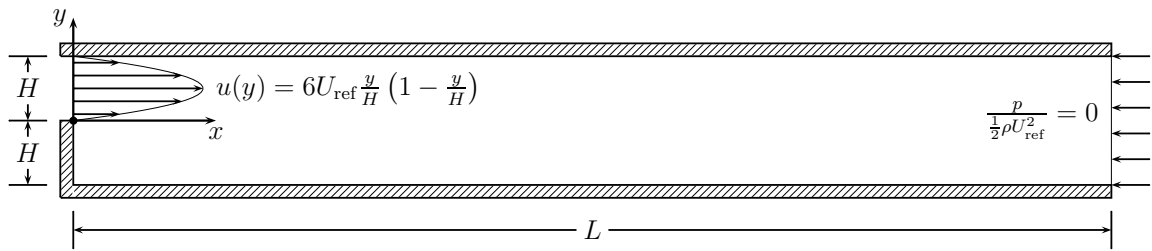}
	\caption{Schematic of the backward facing step flow at ${\rm Re}=\rho U_{\rm ref}(2H)/\mu$}
	\label{fig:7}
\end{figure}

The second test case that we use to examine the performance of the proposed truncation error
estimation scheme  is the laminar flow over a confined backward step~\cite{barton:1995}, 
shown in Figure~\ref{fig:7}. In this work, the Reynolds number based on the total
height of the domain, $2H$, and the average inlet velocity, $U_{\rm ref}$ is 400 and
$L=10H$. The inlet velocity profile is profile is parabolic and the outlet pressure 
is set to zero.

The main difference between the backward facing step flow and the cavity flow of the
previous section is the role of the pressure gradient. In the backward facing step
flow, the pressure gradient is mild everywhere in the domain except near the sharp 
corner of the step. Therefore it is easier to examine the effect of the viscous term
on the performance of the proposed truncation error estimation method.

To obtain the basic second-order solution, we use a uniform unstructured 
triangular mesh with 5120 control volumes. To obtain the solution truncation error
on the original mesh, we subdivide each face into 16 smaller faces. A mesh refinement
analysis confirm that the truncation error calculated based on conservation operator on this
mesh is adequately accurate with an error level smaller than one percent.

\begin{figure}
  \begin{center}
    \subfigure[Actual mass truncation error]{
      \label{fig:8-a}
      \includegraphics[width=\textwidth]{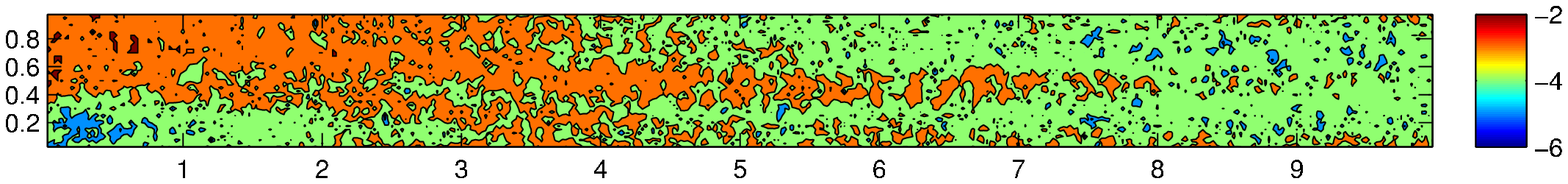}
    }
    \hfill
    \subfigure[Estimated mass truncation error]{
      \label{fig:8-b}
      \includegraphics[width=\textwidth]{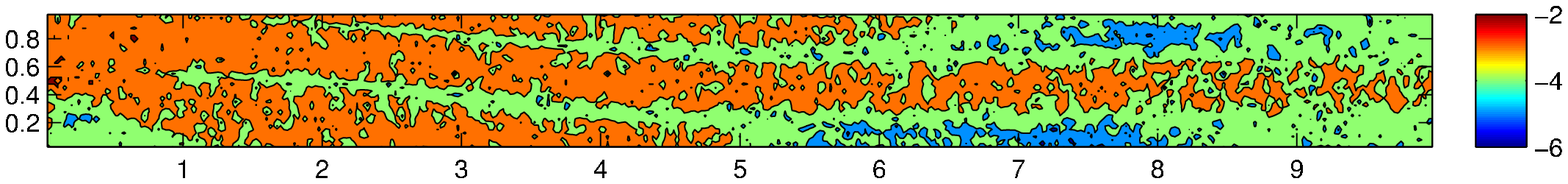}
    }
    \subfigure[Actual $x$-momentum truncation error]{
      \label{fig:8-c}
      \includegraphics[width=\textwidth]{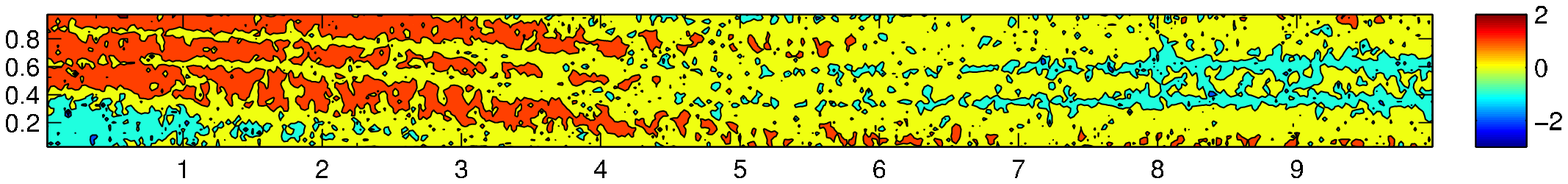}
    }
    \hfill
    \subfigure[Estimated $x$-momentum truncation error]{
      \label{fig:8-d}
      \includegraphics[width=\textwidth]{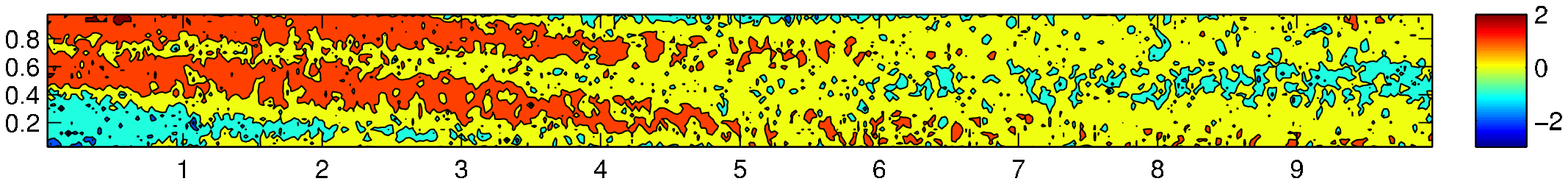}
    }
    \subfigure[Actual $y$-momentum truncation error]{
      \label{fig:8-e}
      \includegraphics[width=\textwidth]{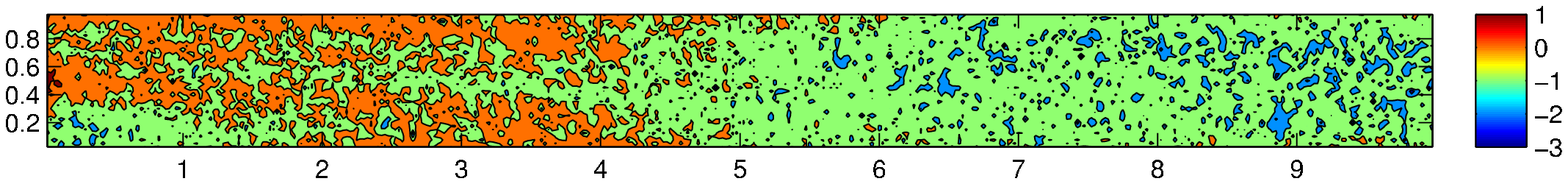}
    }
    \hfill
    \subfigure[Estimated $y$-momentum truncation error]{
      \label{fig:8-f}
      \includegraphics[width=\textwidth]{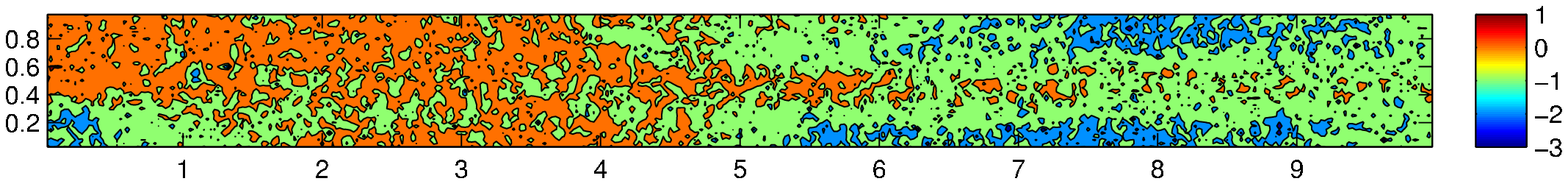}
    }
    \caption{Comparison between the actual and estimated truncation error distributions for
    the backward-facing step flow at ${\rm Re}=400$ on an isotropic triangular mesh.}
    \label{fig:8}
  \end{center}
\end{figure}

Figure~\ref{fig:8} shows the final results for the volume-based truncation error of the mass and 
momentum equations. As seen, the overall trends of the estimated truncation errors are similar
to those of actual truncation errors. However, there are certain discrepancies between the
estimated and the actual truncation error. For example the results for the mass truncation error, 
Figures~\ref{fig:8-a} and~\ref{fig:8-b} show that the proposed scheme overestimates
the truncation error in the separation bubble behind the step. In contrast, the proposed scheme
underestimates the truncation error near the outlet boundary in the vicinity of walls. Similar
discrepancies can also be found in the $x$ and $y$ momentum truncation error. The main reason
for this discrepancies is the way that we estimated the viscous force error across a face.

Equation~\eqref{eq:26} shows the first neglected term in the discretization of
the face viscous force, which is first order. A brief examination of this term
reveals that on a uniform mesh, the vector ${\bf r}_1 +{\bf r}_2=0$ (see Figure~\ref{fig:2}).
Therefore the proposed truncation error estimation scheme fails to resolve 
face viscous force errors on uniform meshes. To fix this problem, higher order
terms must be taken into account. Unfortunately this leads to the appearance 
of third order derivatives of the solution variables, which are not easy to estimate
for a second-orders solution.

\section{Conclusion}
This work proposed a simple truncation error estimation scheme in the context of the
finite volume method. In this context, the solution truncation error in each control
volume is the sum of flow errors across the faces of the control volume. To 
calculate the face flow errors, the first neglected terms in the discretization
of the mass and momentum flows across faces were used. Then these terms were
assembled to obtain the solution truncation error. The numerical experiments with
the lid-driven cavity and backward facing step flow showed that the proposed
scheme successfully estimated the overall trend of truncation error distribution.
However in the backward facing step flow, certain discrepancies were observed
between the estimated and the actual truncation error due to the lack of accurate in
resolving viscous features in the flow. The main advantage of the proposed
scheme was its simplicity in the sense that only second-order derivatives
need to be reconstructed. However in the cases that the viscous term in 
the momentum equation is sole source of error, the proposed truncation error
estimation scheme may become less accurate.

\bibliographystyle{plain} 
\bibliography{ref} 

\end{document}